\documentclass{elsart}

\usepackage{graphicx}

\usepackage{amssymb}

\begin{document}

\begin{frontmatter}

\title{Performance of the triple-GEM detector with optimized 2-D readout
        in high intensity hadron beam.}

\author{        A.Bondar, A.Buzulutskov, L.Shekhtman, A.Sokolov,  A.Vasiljev}




\address{ Budker Institute of Nuclear Physics, 630090 Novosibirsk, Russian Federation }

\begin{abstract}
Multiple-GEM detectors are considered to be good candidates 
for tracking devices in experiments with high hadronic background.
We present the results of the triple-GEM detectors beam test in a high intensity pion
beam. The detectors had an optimized  two-dimensional readout with minimized strip 
capacitance. Such optimization permitted the starting point of the efficiency plateau
down to a gain of 5000. The probability of GEM discharges induced by heavily ionizing particles 
has been measured as a function of gain: at a gain of 20000 it amounts to $10^{-11}$ 
per incident particle. Such a value will ensure safe operation of the detector
in the conditions of forward region of the LHC experiments.

\end{abstract}


\end{frontmatter}

\section{Introduction}

 Micro-pattern gas technologies have been considered as a good candidates for
inner tracking systems of LHC experiments. A major problem of gas micro-pattern devices
appeared to be heavily ionizing particles that are produced in nuclear interactions
of hadrons with the material of the detectors. High and dense ionization produced in a sensitive
region of the detector can provoke sparking with subsequent deterioration of its properties
(gain and efficiency) and possible destruction. It was shown that introducing several 
gas amplification stages one can significantly improve performance of micropattern devices 
in this respect [1]. The most convenient and safe way of producing such multistage gas
amlification was application of Gas Electron Multiplier (GEM)[2]. The last development in this
direction, the Triple-GEM detector, utilizing 3 consequtive GEM foils was shown to be the
most safe with respect to sparking in hadronic environment [3,4]. 

 Triple-GEM detectors have been developed as a possible technology for the inner tracking
in LHCb experiment [3,5]. Simulation of the LHCb interaction region, including the beam vacuum 
pipe, were performed to estimate the particle fluxes [6], using an average nominal luminosity
of 2*$10^{32}$cm$^{-2}$s$^{-1}$. The particle fluxes and its compositions depend strongly on 
the position along the beam axis and on the distance from the beam. Maximum charged hadron
rate is expected to be $8*10^3mm^{-2}s^{-1}$ and total hadron rate per station reaches 100 MHz. 
It was shown that the GEM foil can withstand more than 100 sparks per cm$^2$ without permanent damage[7].
If we require that sparking rate is limited to 1 per $10^3$s per tracking plane and take into account
local and total hadronic rate, it follows that the discharge probability per incident particle 
must not exceed $10^{-11}$. 

  In this paper we describe the Triple-GEM detector with optimized readout board. New readout
allowed to reduce significantly interstrip capacitance and thus improve signal to noise ratio
of the detector. The results of test in high intensity pion beam demonstrate considerable improvement
in the performance related to discharges due to heavily ionizing particles.  
  
\section{Detector design and experimental set-up.}

 The Triple-GEM detector consists of a primary ionization gap, 3 consequtive GEM foils separated by
transfer gaps and a readout Printed Circuit Board (PCB), separated from the bottom GEM by an induction
gap. The design and general properties of multi-GEM detectors were discussed in details earlier [3,4,5,8].
Here we will concentrate mainly on the design of readout PCB and its influence on operation of the detector.
In LHCb each tracking plane of the Inner tracking system has to provide position in horizontal direction 
with a precision of better than $\sim$200$\mu$ to keep momentum resolution below 0.5\%. In the vertical 
direction, position accuracy can be 10 times worse as it is needed only for pattern recognition. Thus a 
stereo readout is foreseen with an angle of 0.1 rad. 
   
The first measurements performed with large size prototype with small angle stereo readout [3] showed that 
the basic limitation on signal to noise ratio was high strip capacitance of the PCB. In the case
of [3] it was about 100 pF for 30 cm long strips. 

The PCB for such a detector is produced in two layers separated one from the other by a $50\mu$ kapton foil.
Between metal strips of the top layer the kapton layer is etched out and the metal of bottom layer
is opened. We propose to make zero degree strips at the top layer and small angle stereo strips and the 
bottom layer. The bottom strips have to be made in short sections parallel to the top ones with narrow 
``bridges'' connecting the sections belonging to one bottom strip. Schematic drawing of this layout is
shown in fig.1. 
\begin{figure}[htbp]
\includegraphics[width=13cm, height=13cm]{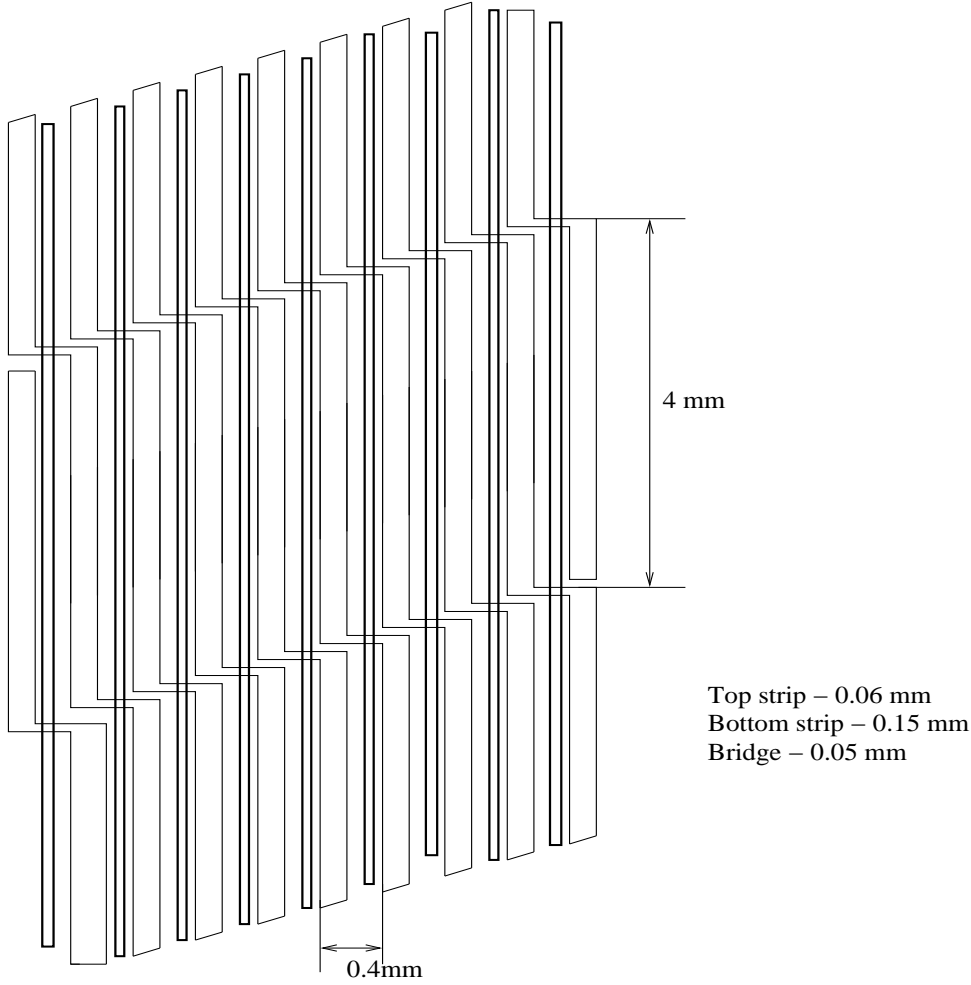}
\caption{Schematic view of the PCB layout.}
\end{figure}
 
 In such a layout the area of overlapping regions between top and bottom metal layers is reduced, thus 
minimizing the interstrip capacitance. Moreover, as the bottom and top strips are parralel, the sharing of 
induced charge between  will be constant, and will not depend on the position along 
the strip. This is not the case for the layout in [3] where top strips were not parallel to those at
the bottom. 

 In order to reduce further the interstrip capacitance the strips have to be made as narrow as possible. 
A natural limitation for the width is the feature size of the technology, that is about $50\mu$. 
Thus the width of the top strip was chosen as $60\mu$. The ratio of the strip widths is determined by the 
requirement to have the same signal to be induced on top and bottom strips. As top and bottom strips 
are separated by $50\mu$ and  signal is induced by charge moving from the last GEM through the
induction gap, bottom strips have to be wider than the top ones. We used MAXWELL package to calculate 
the necessary ratio of widths and strip capacitances. The thickness of copper layers was assumed to be $5\mu$.
The kapton layer was $50\mu$ thick and the width of kapton strip at the bottom was taken $10\mu$ larger than 
at the top (at the contact with top strip). We also introduced in the model thick (1mm) epoxy substrate
below the bottom strips. The calculations were made for a strip pitch of $400\mu$.

 In the calculations we required that the ratio of the signals induced on the bottom strips with respect
to the top ones is equal to 1.2, in order to compensate for the larger capacitance of the bottom strips.
The results of the calculations are shown in table 1.

 In order to achieve the sharing of induced signals as indicated above the width of the bottom strips has to
be $150\mu$. The values of calculated strip capacitance per cm of length are shown in the first row of
Table 1. The capacitance of the bottom strip is almost 2 times higher than that of top strip. However
we can see that for 30cm long strip total capacitance will be below 20pF.  In the second row
of the table
the results of measurements are shown. A $10*19 cm^2$ prototype PCB was used for the measurments with all 
strips grounded around the one which capacitance was measured. Experimental results are higher than 
calculated ones probably due to the differences in the particular shape of metal and kapton strip edges.

\begin{table}[htbp]
\begin{tabular}{c|c|c|c}\hline
 layer & strip width &  pF/cm (calculation) & pF/cm (measurement) \\
\hline
 bottom & $150\mu$ & 0.62 & 0.73 \\
 top & $60\mu$ & 0.32 & 0.54 \\
\hline
\end{tabular}
\vspace{1cm}
\caption{Strip widths and capacitances for the optimized PCB.}
\end{table}

 The prototype boards were produced with parameters shown in table 1 and size of sensitive
area of $10*10 cm^2$. Top and bottom strips were extended from both sides forming very long bonding pads
with a length of 4.5 cm and effective pitch of $200\mu$. Such long pads were made to have more freedom 
in connecting electronics, and simulate 
the case of a real strip length having smaller sensitive area. 

 The Triple-GEM detectors were assembled with prototype boards and GEM foils with $10*10 cm^2$ sensitive 
area. All transfer gaps were 1mm and induction gaps  2 mm. In the transfer and induction gaps
we put bent strips of mylar 1mm and 2mm wide respectively that were attached to the frame at both ends.
These strips served as spacers keeping GEM foil at a precise distance from the PCB and from each other.
The drift gap was kept as 3mm. 

 Two detectors were equipped with PREMUX hybrid with 512 channels each. Detailed description of PREMUX
can be found in [9]. Each channel of PREMUX contains low-noise preamplifier, shaper and analogue buffer.
Analog buffers can be readout sequentially through 1MHz multiplexer. Top and bottom strips were 
connected to the channels of PREMUX in series, i.e. each "stereo" channel was followed by a "straight"
channel. The separation of signals from  "stereo" and "straight" strips has been made off-line 
during data analysis.

 In order to study the performance of the Triple-GEM detectors in high intensity hadron beams,
assembled devices were exposed to the beam of 350 MeV pions at the proton cyclotron in Paul 
Scherrer Institute (PSI, Villigen, Switzerland). The beam was tuned to have maximum intensity
of $\sim10^4 mm^{-2}s^{-1}$. The width of the beam was $\sim9$ cm (FWHM) and the height was
$\sim5$ cm. Total beam intensity within the area of the detectors ($10*10 cm^2$) was $6*10^7 s^{-1}$. 
With this beam intensity we could measure discharge probablity below $10^{-11}$ with reasonable 
statistical significance in several hours. 

 A schematic layout of the set-up at the beam is shown in fig.2. The two detectors were attached to the 
bench as close as possible to each other between two scintillating counters. The counters were
used for the measurements of gain and efficiency at low intensity. For these measurements
the beam intensity was reduced down to $\sim100-1000$Hz. Coincidence of scintillating counters
was used as a trigger that produced sample-and-hold signal for PREMUX. Analog signals from each
strip of the detectors were stored in the buffers of PREMUX, readout and digitized by a sampling ADC
and then stored in the computer. More detailed description of the Data Aqcuisition system (DAQ) for 
PREMUX can be found in [10]. 

\begin{figure}[htbp]
\includegraphics[width=10cm, height=6cm]{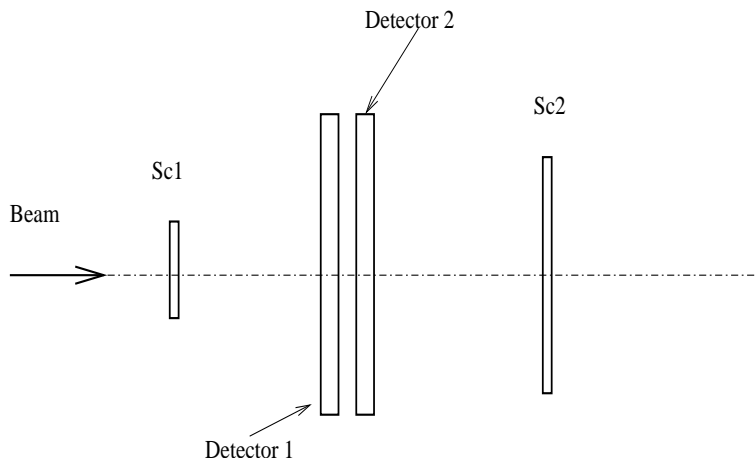}
\caption{Schematic view of the set-up.}
\end{figure}

 The correspondence between signal after the ADC and the input charge was found before the measurements
using built-in calibration capacitances. With this electronic calibration we could determine
input charge and gain of the detectors.  

 For counting of the discharges we measured the current of the drift cathode. In the presence of full 
beam intensity this current was higher than 1nA when the gain was above $10^3$. However when a
discharge occured this current dropped to zero and then restored according to the recovery of GEM
voltage. Using a current meter with analog output proportional to the input current, we assembled 
a simple set-up including an  amplifier, discriminator and scaler. The threshold of this set-up was
tuned in such a way that all the discharges were counted. 

\section{Results and discussion.}
 
 The main goal of this work was to study the dependence of discharge probability on gas gain in the
detector and compare it with the efficiency versus gain performance. The beginning of the efficiency plateau
is determined by the primary ionisation deposited in the drift gap and noise of the electronics. 
In fig.3 noise values (sigma of the gaussian fit) are shown for
all channels of one of the detectors, for straight strips (top figure) and stereo strips (bottom
figure).
We can see that the noise for stereo strips is higher ($\sim9$ ADC bins) than for straight strips
($\sim6$ ADC bins).
From electronic calibration we knew that 1 ADC bin corresponded to 125 electrons. Thus the noise value
for stereo channels was $\sim1200$e and for straight channels it was $\sim800$ e. At the left side of
the bottom 
figure we can see some reduction of the noise as stereo strips become shorter reaching the side edge of the structure 
rather than the opposite edge. 

\begin{figure}[htbp]
\includegraphics[width=13cm, height=13cm]{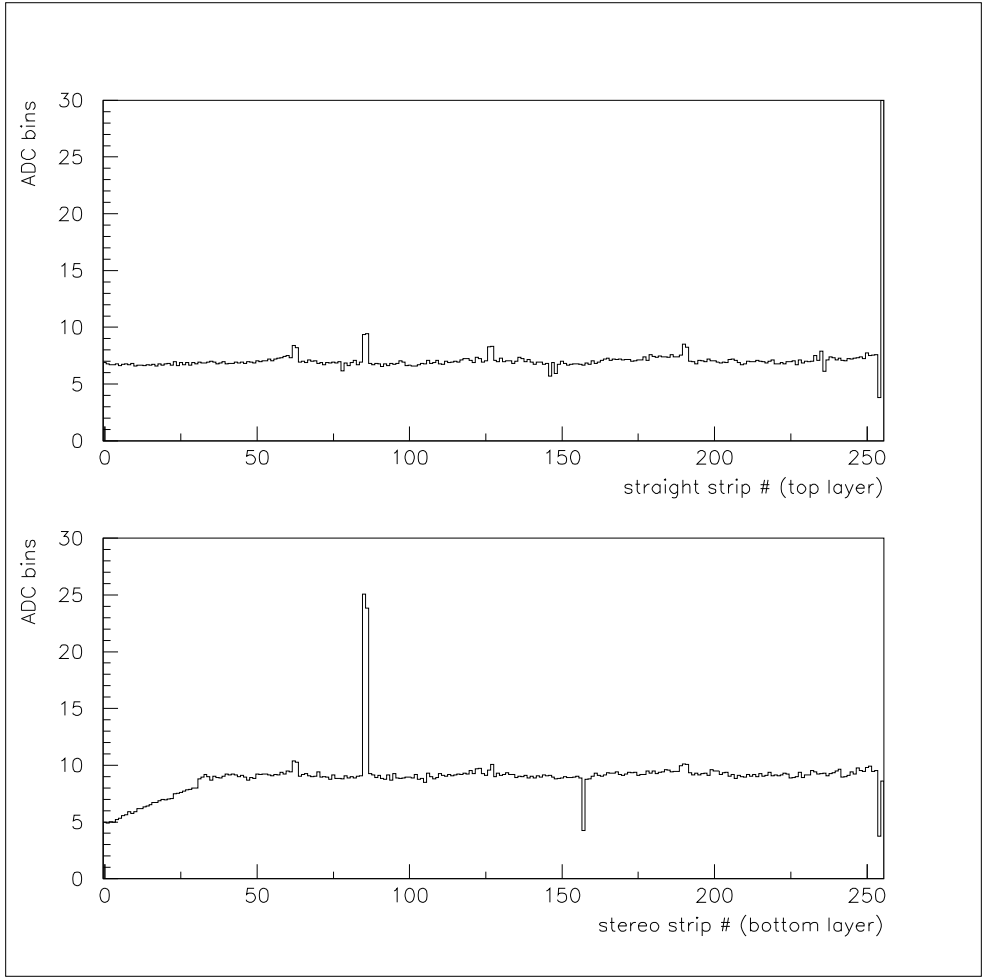}
\caption{Noise values (sigma of gaussian fit) for all channels of the detector.}
\end{figure}

  In fig.4 typical signals from charged particle are shown. Top and bottom figures correspond to
straight and
stereo strips respectively. In both layers two groups of channels(clusters) have signals. Higher signal
obviously corresponds to the main charge induced by the incident particle, while the smaller signal is a
pick-up from the opposite layer. This conclusion is confirmed by coincidence of the smaller cluster position
with the main one from the opposite layer. Such a strong pick-up is determined by very long extensions of 
the strips outside sensitive area where strips from both layers go parellel to each other. In the final detector
such a design should be avoided and strips at the regions of fan-outs and bonding pads have to be as short 
as possible. 

\begin{figure}[htbp]
\includegraphics[width=13cm, height=13cm]{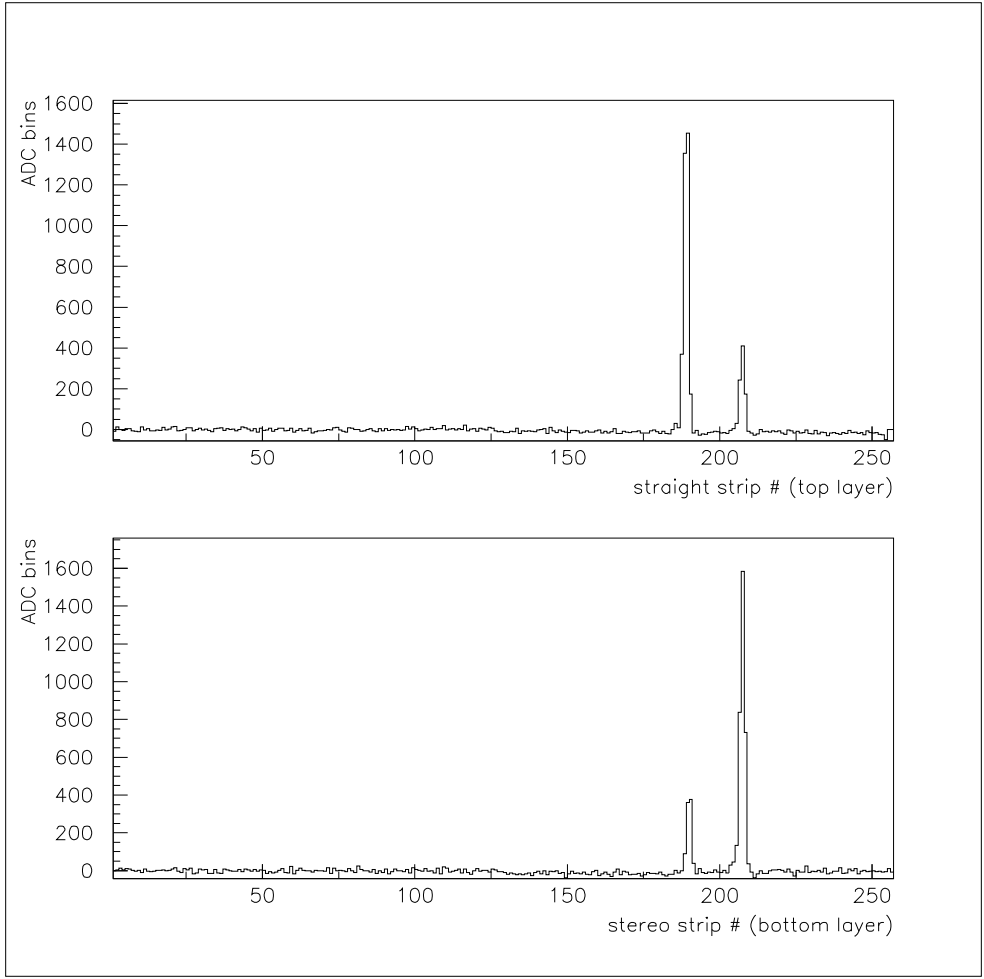}
\caption{Typical signal from charged particle. Response of top and bottom layers is shown.}
\end{figure}

 In order to find the cluster corresponding to an incident particle, first, all channels were
sorted into straight and stereo ones. Then those channels that exceeded a certain threshold (usually
2-3 sigma noise) were found. Signals were summed up within continous groups of such channels (clusters).
We took always only the cluster with the highest charge for further analysis. An example of cluster 
charge distribution is shown in fig.5. We can see clear separation of the main
part of the signal from noise peak. Here the cluster signals from both layers were summed together.

\begin{figure}[htbp]
\includegraphics[width=13cm, height=13cm]{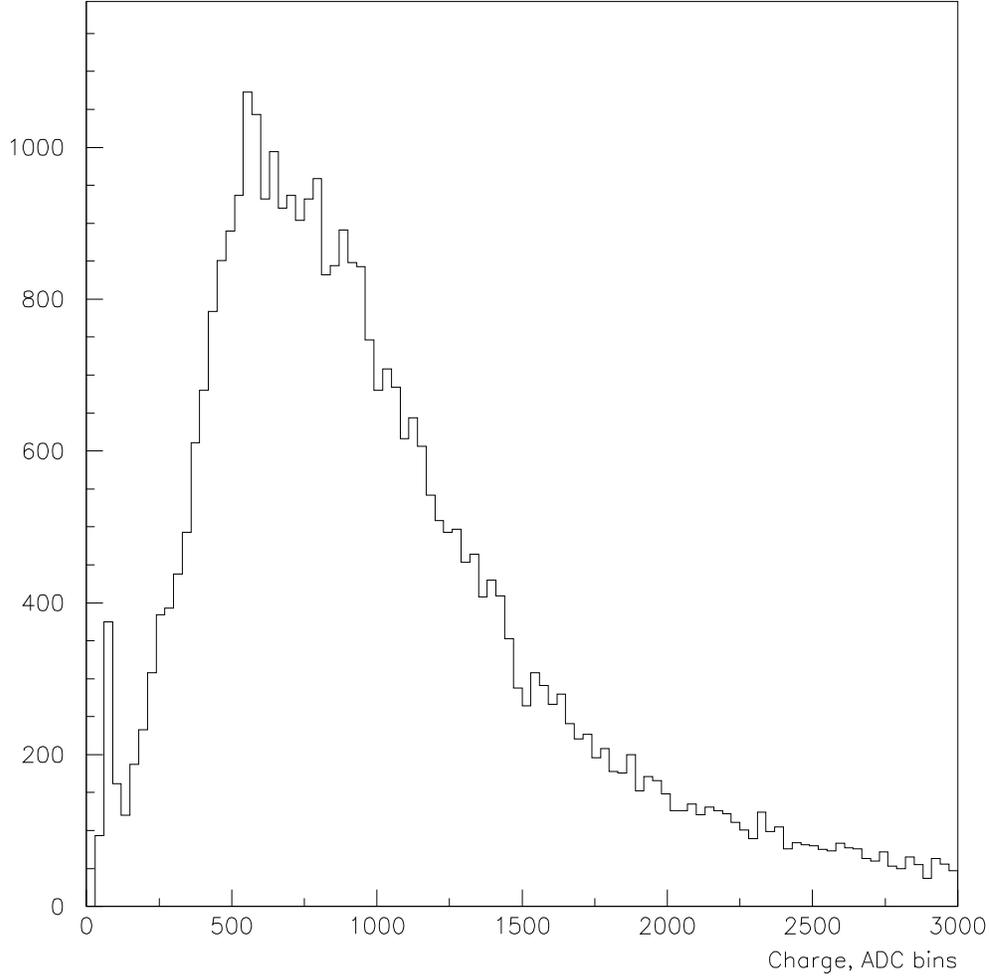}
\caption{Typical cluster charge distribution.}
\end{figure}
 
 The relationship between signals in top and bottom layers is shown in fig.6. Here the correlation between
cluster charge in top and bottom layers is plotted. We see that these signals are almost equal to each other.
Thus the result of this measurement is lower by 20\% than the value expected from the calculations. 
Signals that are seen at the sides of the figure can appear when the main cluster 
is lost and the ``pick-up'' signal is taken instead. Losses of signal might happen due to two noisy 
and two broken channels in the bottom layer that were excluded from the analysis (see fig.3). 

\begin{figure}[htbp]
\includegraphics[width=13cm, height=13cm]{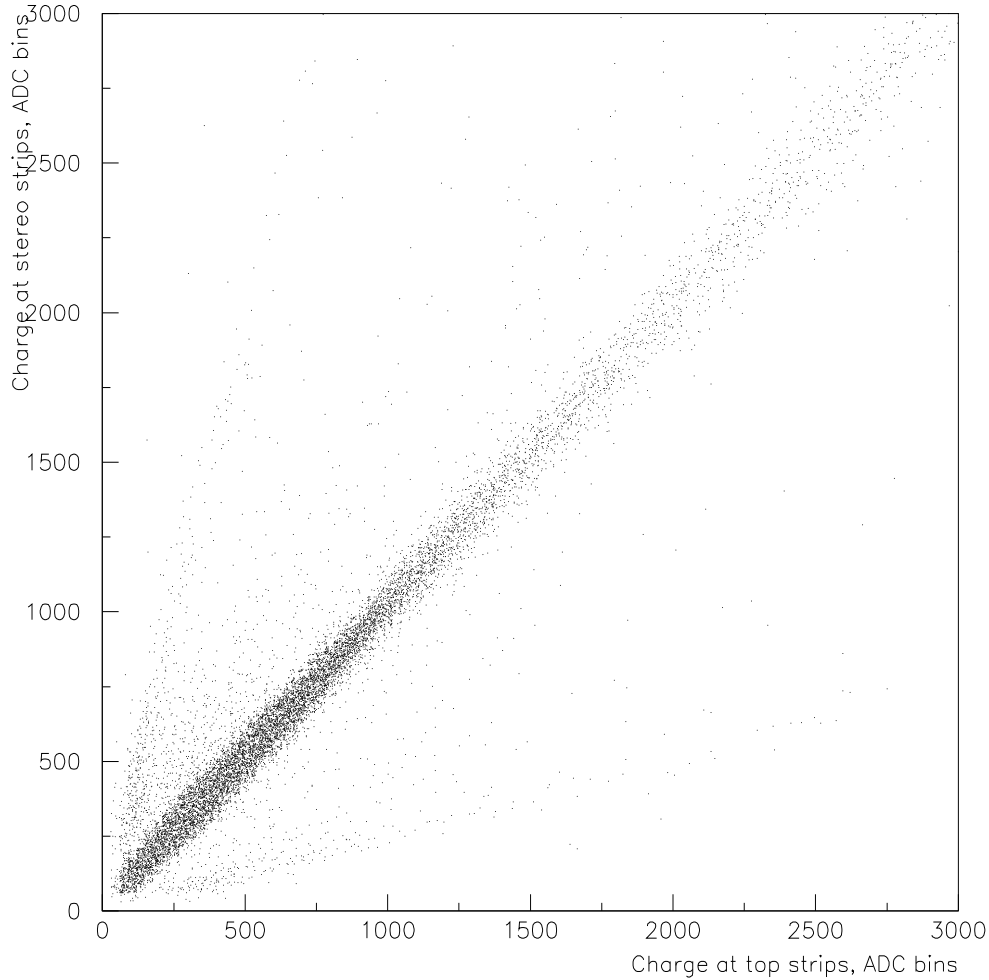}
\caption{Correlation between cluster charge in top and in bottom layers.}
\end{figure}

 The pion beam was tuned in such a way that most of the sensitive area of the detectors was irradiated.
In fig.7 the distribution of reconstructed cluster positions in two dimensions is shown. This distribution
demonstrates roughly the  beam intensity within the area of the detector and scintillator
counters.  Binning in horizontal direction is determined by the spacing
of straight strips (0.4 mm), while that in the vertical direction corresponds to the pitch of stereo
structure (4 mm), 

\begin{figure}[htbp]
\includegraphics[width=13cm, height=13cm]{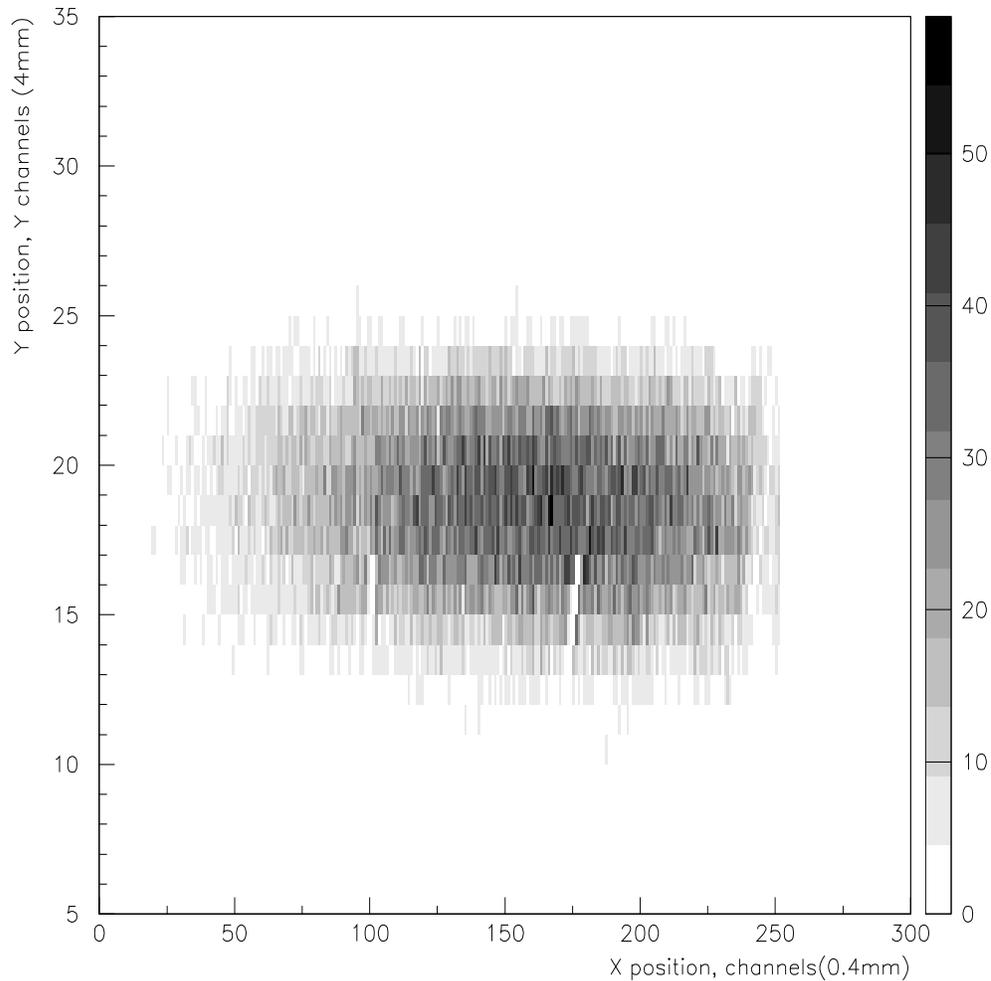}
\caption{Beam profile measured at low intensity at detector 2.}
\end{figure}

 Gas gain of the detectors was determined as the ratio of average cluster charge over
the average number of primary electrons released by minimum ionising particle in the 3 mm
gap. The latter is equal approximately to 30 electrons for the gas mixture used ($Ar-CO_2$,
70-30). In fig.8 the dependence of gain on GEM voltage is shown for both detectors. Voltage differences
from top to bottom electrodes in all GEMs  were kept the same. The values of field in transfer, induction and
drift gaps are indicated in the figure. The difference in gain at the same GEM voltage in two detectors 
can be explained  by a limited precision of spacing of the GEM foils, accuracy of resistors in 
the resistive network and possible difference in hole diameter in different GEMs.  

\begin{figure}[htbp]
\includegraphics[width=13cm, height=13cm]{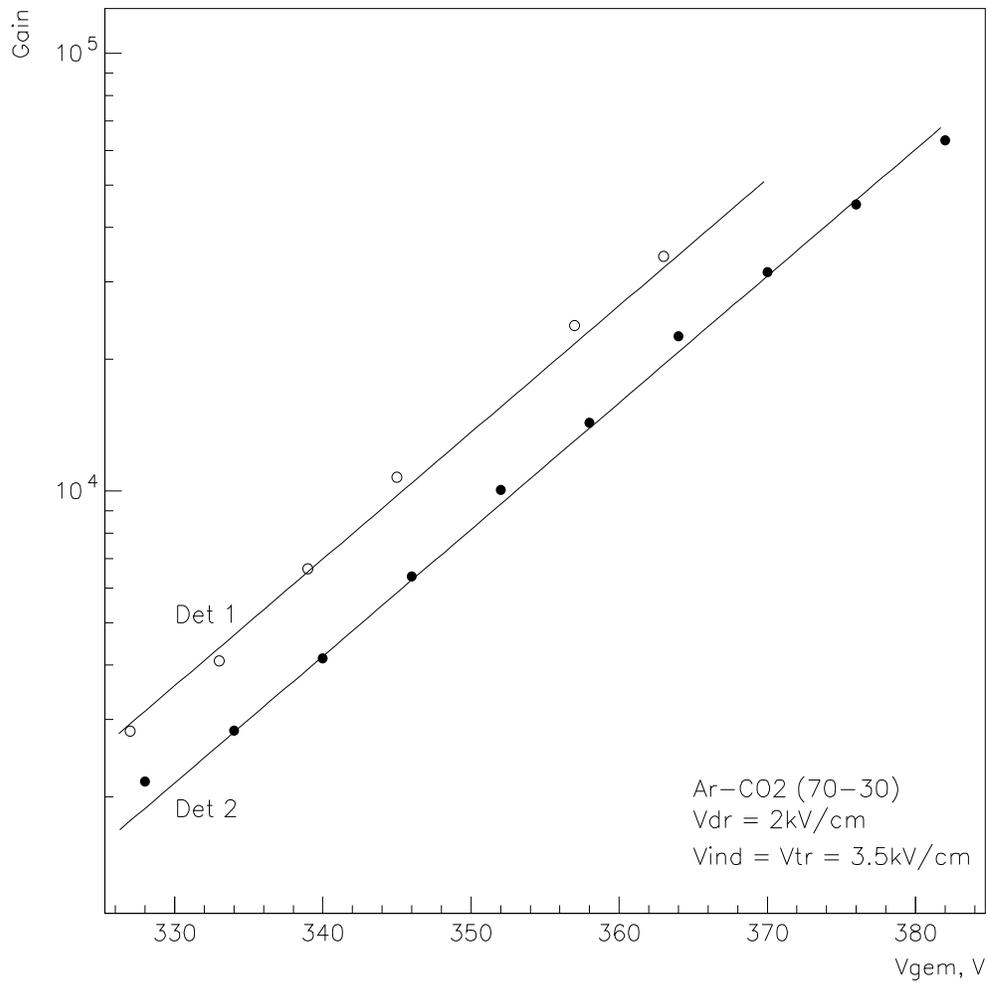}
\caption{Dependence of gain on GEM voltage in two detectors.}
\end{figure}

 As we did not have any tracking devices apart from the detectors under study, we used one detector to 
determine the efficiency of the other. Efficiency of detector 2 was defined as the ratio of the number of cases
when signals were found in both detectors to the total number of cases when detector 1 had a signal. 
With this definition of the efficiency we could not get values close to 100\% because the detectors
were not precisely aligned. Some tracks that were detected by detector 1 did not pass through detector
2. This problem however did not prevent to define the starting point of efficiency plateau. The dependence 
of efficiency on the gain together with the discharge probability is shown in fig.9. The values of efficiency are
indicated at the left scale and the values of discharge probability at the right. The beginning of efficiency
plateau is clearly determined at a gain of $\sim5000$ while the value of probability equal to $10^{-11}$ is 
achieved at a gain of $\sim20000$. There is a margin of a factor 4 between the beginning of efficiency plateau and 
discharge threshold  where the detector can be operated safely.

\begin{figure}[htbp]
\includegraphics[width=13cm, height=13cm]{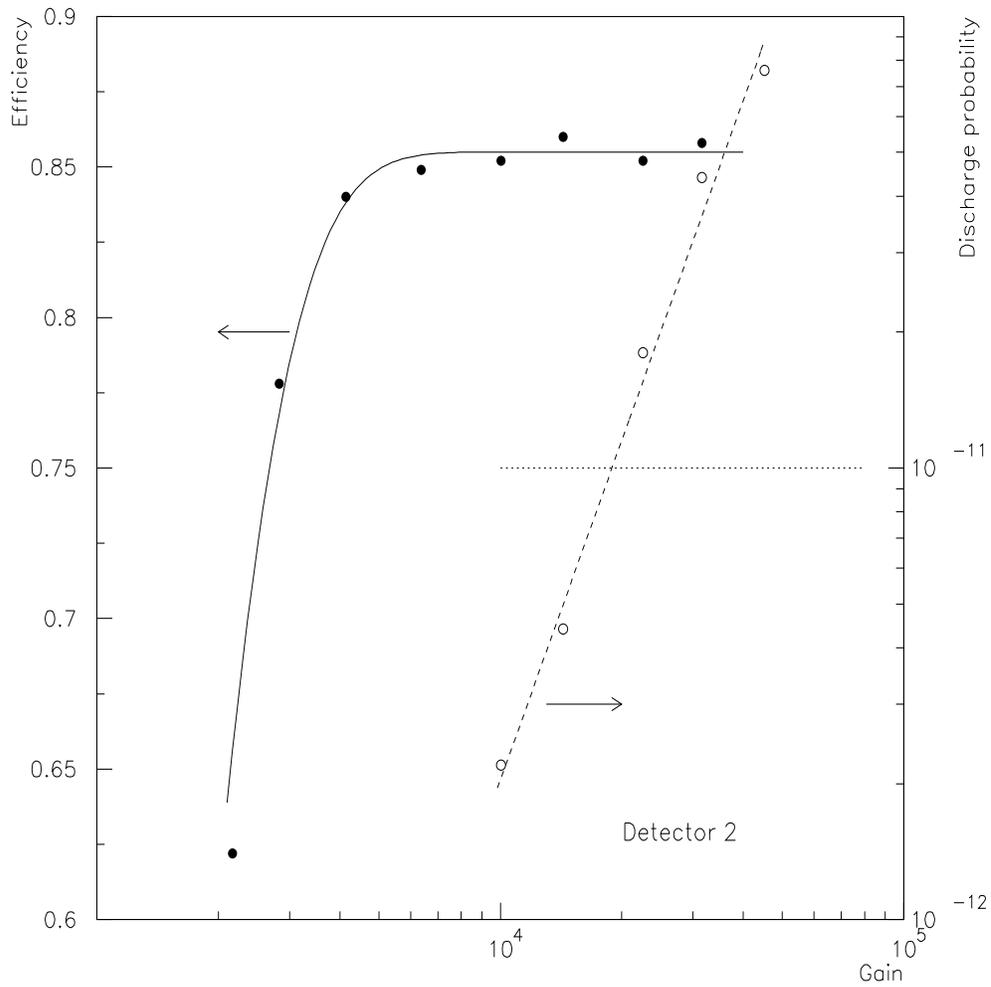}
\caption{Dependence of efficiency (left scale) and discharge probability per incident particle(right scale) on gain.
Dashed line is indicating the level of $10^{-11}$ considered as a safe limit.}
\end{figure}
 
\section{Conclusions}

 Safe operation of the Triple-GEM detector was demostrated with a margin of a factor 4 between the beginning of efficiency 
plateau and gain where the discharge probability per incident particle exceeded $10^{-11}$. The detector had 
a sensitive area of $10*10 cm^2$ but the effective strip length was 19 cm. 
This result was possible due to an optimized design of 
the readout PCB that allowed significant reduction of the strip capacitance down 
to 0.5-0.7 pF/cm. The detector still 
suffers from pick-up between the layers of PCB. In the final design  the regions of fan-outs and bonding pads
will be made as short as possible and this effect will be significantly suppressed.
   
\section{Acknowledgements}

 The authors would like to thank very much J.-P.Perroud, P.Sievers, M.Ziegler 
and U.Straumann for significant help
during the test period in PSI, L.Ropelewski and F.Sauli for assistance 
in preparation at CERN, D.Renker for 
help at the experimental area in PSI.

\end{document}